\def\year{2022}\relax
\documentclass[letterpaper]{article} % DO NOT CHANGE THIS
\usepackage{aaai22}  % DO NOT CHANGE THIS
\usepackage{times}  % DO NOT CHANGE THIS
\usepackage{helvet}  % DO NOT CHANGE THIS
\usepackage{courier}  % DO NOT CHANGE THIS
\usepackage[hyphens]{url}  % DO NOT CHANGE THIS
\usepackage{graphicx} % DO NOT CHANGE THIS
\usepackage{natbib}  % DO NOT CHANGE THIS AND DO NOT ADD ANY OPTIONS TO IT
\usepackage{caption} % DO NOT CHANGE THIS AND DO NOT ADD ANY OPTIONS TO IT
\DeclareCaptionStyle{ruled}{labelfont=normalfont,labelsep=colon,strut=off} % DO NOT CHANGE THIS
\usepackage[ruled,linesnumbered]{algorithm2e}
\usepackage{appendix}
\usepackage{newfloat}
\usepackage{listings}

\usepackage{multirow}
\usepackage{subcaption}
\usepackage{xcolor}
\usepackage{amsmath}
\DeclareMathOperator*{\argmax}{arg\,max}
\usepackage{enumitem}

\usepackage{amsfonts}
\usepackage{bm}
\title{NSGZero: Efficiently Learning Non-Exploitable Policy in Large-Scale Network Security Games with Neural Monte Carlo Tree Search}
\author {
    Wanqi Xue, 
    Bo An, 
    Chai Kiat Yeo 
}
\affiliations {
     School of Computer Science and Engineering, Nanyang Technological University, Singapore\\
    wanqi001@e.ntu.edu.sg, boan@ntu.edu.sg, asckyeo@ntu.edu.sg
}

\usepackage{bibentry}
\begin{document}
\maketitle
\begin{abstract}
% Network Security Games (NSGs) permits the modelling of resource deployments for securing critical targets in networks. 
How resources are deployed to secure critical targets in networks can be modelled by Network Security Games (NSGs). While recent advances in deep learning (DL) provide a powerful approach to dealing with large-scale NSGs, DL methods such as NSG-NFSP suffer from the problem of data inefficiency. Furthermore,  due to centralized control, they cannot scale to scenarios with a large number of resources. In this paper, we propose a novel DL-based method, NSGZero, to learn a non-exploitable policy in NSGs. NSGZero improves data efficiency by performing planning with neural Monte Carlo Tree Search (MCTS).
Our main contributions are threefold. First, we design deep neural networks (DNNs) to perform neural MCTS in NSGs. Second, we enable neural MCTS with decentralized control, making NSGZero applicable to NSGs with many resources. Third, we provide an efficient learning paradigm, to achieve joint training of the DNNs in NSGZero. Compared to state-of-the-art algorithms, our method achieves significantly better data efficiency and scalability.
\end{abstract}

\section{Introduction}
Network Security Games (NSGs) have been used to model the problem of deploying resources to protect important targets in networks~\cite{okamoto2012solving,wang2016computing,wang2020scalable}. Many real-world security problems, including infrastructure protection~\cite{Jain11}, wildlife conservation~\cite{fang2015security} and traffic enforcement~\cite{yz17,yz19,xue2021solving}, can be boiled down to NSGs. In NSGs, a defender, controlling several security resources, interacts adversarially with an attacker. The objective of the attacker is to take a path from the starting point to a target without being intercepted by the defender. The defender's goal is to develop a resource allocation policy to interdict the attacker. Traditionally, mathematical programming-based approaches, e.g., the incremental strategy generation algorithm~\cite{doubleoracle}, are proposed to compute the optimal policy for the defender. However, limited scalability prevents programming-based approaches from being applicable to complex and real-world NSGs~\cite{xue2021solving}.

Recent advances in deep learning (DL) and reinforcement learning have led to remarkable progress in playing complex games, with successful applications surpassing human performance in Go~\cite{silver2016mastering}, chess~\cite{alphazero}, poker~\cite{deepstack} and video games~\cite{dqn,alphastar,atari}. Deep neural networks (DNNs), with strong representation ability, are able to capture underlying structures of enormous game state spaces when empowered with sufficient computational resources. NSG-NFSP~\cite{xue2021solving}, a DL-based approach for solving NSGs,
integrates representation learning with the framework of NFSP~\cite{nfsp} and enables NFSP with high-level actions to achieve efficient exploration in large-scale NSGs.

Although DL provides a powerful approach to dealing with large-scale NSGs, existing approaches, e.g., NSG-NFSP, have largely neglected intrinsic properties of the environment's dynamics. Concretely, when a player in NSGs selects an action (the node it will move to), the next state of the game can be partially determined~\footnote{The next state cannot be fully determined because players in NSGs act simultaneously. No player can know what action its opponent will take in the next step.}, because the next location of the player can be inferred from the chosen action. 
However, in NSG-NFSP, it samples many rounds of the game and uses Monte Carlo method to estimate the distribution of the next state from scratch, leading to poor data efficiency. Another problem is that the centralized control of security resources in NSG-NFSP will inevitably result in combinatorial explosion in action space~\cite{oliehoek2008optimal}, making the algorithm unsuitable for handling scenarios where there are many resources. 

In this work, we propose a DL-based method, NSGZero, for efficiently approaching a non-exploitable defender policy in NSGs. NSGZero improves data  efficiency by modeling the dynamics of NSGs and performing planning with neural Monte Carlo Tree Search (MCTS)~\cite{coulom2006efficient}. 
Our key contributions are in three aspects. First, we design three DNNs, namely the dynamics network, the value network and the prior network, to model environment dynamics, predict state values, and guide exploration, respectively, which unlock the use of efficient MCTS in NSGs.
Second, we enable decentralized control in neural MCTS, to improve the  scalability of NSGZero and make it applicable to NSGs with a large number of security resources.
Third, we design an effective learning paradigm to joint train the DNNs in NSGZero. Experimental results show that, compared to state-of-the-art algorithms, NSGZero achieves significantly better data efficiency and scalability.

\section{Related Work}
% In this section, we discuss some recent works related to our method. We first introduce works on integrating MCTS with neural network function approximation. Next, we discuss recent attempts on applying DL for solving security games.

\textbf{Neural MCTS.} Monte Carlo Tree Search~\cite{coulom2006efficient} is a planning method which explores possible future states and actions by querying a simulator or model of the environment. 
At each decision point, MCTS repeatedly performs multiple simulations, to evaluate the probability of choosing each available action. There have been many attempts to combine MCTS with neural network function approximations to solve complex games, notably the AlphaGo series. AlphaGo~\cite{silver2016mastering}, the first algorithm that defeats human professional players in the full-sized game of Go, conducts lookahead searches by
using a policy network to narrow down the decision to high-probability moves and using a value network to evaluate state values in the search tree. AlphaGo Zero~\cite{silver2017mastering} achieves superhuman performance purely from random initialization, without any supervision or use of expert data. AlphaZero~\cite{alphazero} generalizes its predecessor into a single algorithm which can master many challenging domains, including chess, shogi and Go. For the aforementioned algorithms, they assume complete access to the rules of the game. MuZero~\cite{schrittwieser2020mastering} lifts this restriction and uses neural networks to approximate the transition function and the reward function of the environment. 
% In NSGZero, although we allow players to infer state transitions, available actions, and episode termination, the next state of a player cannot be fully determined by these game rules due to simultaneous movement. We use a dynamics network, together with information about the rules of the game, to predict the next state. 
Recently, Sampled MuZero~\cite{pmlr-v139-hubert21a} extends MuZero to complex action space by performing policy improvement and evaluation over small subsets of sampled actions. Despite the great breakthroughs, all the previously discussed approaches have focused on controlling a single agent, while whether and how neural MCTS can be applied to multi-agent scenarios like NSGs remains unexamined.

\textbf{DL for Security Games.} Applying DL to solve security games has recently received extensive attention. DeDOL~\cite{greensecurity} computes a patrolling strategy by solving a restricted game and iteratively adding best response strategies to it through deep Q-learning. OptGradFP~\cite{cfsp1} addresses security games with continuous space by policy gradient learning and game theoretic fictitious play. NSG-NFSP~\cite{xue2021solving} integrates representation learning into the framework of NFSP~\cite{nfsp}, to handle complex action spaces. Despite the progress, these methods are model-free, and they suffer from data inefficiency. Moreover, they are not suitable for high-dimensional action spaces. Recently, CFR-MIX~\cite{cfrmix} is proposed to deal with high-dimensional action spaces. However, it requires traversing the entire game tree, which makes it impractical for games with a large branching factor and long time horizon such as NSGs.

\section{Problem Formulation}
Network Security Games (NSGs) are used to describe the problem of deploying resources to protect against an adaptive attacker in networks~\cite{Jain11}.
An NSG can be formulated by a tuple $\langle G, V_s, V_t, V_m, T \rangle$, where $G=(V,E)$, consisting of a set of nodes $V$ and a set of edges $E$, is the graph on which the NSG is played. $V_s \subset V$ is the set of possible starting nodes of the attacker. The target nodes, which represent destinations to be attacked or exits to escape, are denoted by $V_t \subset V$. The defender controls $m=|V_m|$ resources and the resources start from the nodes in $V_m \subset V$. $T$ is the time horizon. The attacker and resources move on graph nodes, and a move is valid if and only if $(v_t, v_{t+1}) \in E$. Let $v^{att}_t$ and $l_t^{def}=\langle v_t^0,\dots, v_t^{m-1} \rangle$ denote the positions of the attacker and the $m$ resources at step $t$ respectively, following former works~\cite{yz19,xue2021solving}. The state of the attacker $s^{att}_t$ is the sequence of the nodes he has visited, i.e., $s_t^{att}=\langle v_0^{att},v_1^{att},\dots, v_t^{att} \rangle$, and the state of the defender $s^{def}_t$ consists of $s^{att}_t$ and the resources' current locations, i.e., $s_t^{def}=\langle s_t^{att},l_t^{def}\rangle$. The defender and the attacker interact sequentially, and their policies $\pi(s_t)=\Delta(\mathcal{A}(s_t))$ are mappings from state to a distribution $\Delta$ over valid moves (legal actions)~\footnote{We omit the superscripts for $\pi$ and $s_t$ because the formulate applies to both the defender and the attacker.}. Here, $\mathcal{A}(s_t)$ is a function which returns the set of legal actions at $s_t$, e.g., $\mathcal{A}(s_t^{att})=\{v_{t+1}^{att}|(v_{t}^{att},v_{t+1}^{att})\in E\}$. For both players, they have free access to $\mathcal{A}(s_t)$. 
An NSG ends when the attacker reaches any of the target nodes within $T$ or is caught~\footnote{The attacker is caught if he and at least one of the resources are in the same node at the same time or the time is up.}. If the attacker is caught, the defender will receive an end-game reward $r^{def}=1$, otherwise, no award will be awarded to her. The game is zero-sum, so $r^{att}=-r^{def}$. A policy is said to be non-exploitable if it achieves the best performance in the worst-case scenario. Therefore, the optimization objective in NSGs is to maximize the worst-case defender reward $\max_{\pi^{def}}\min_{\pi^{att}} \mathbb{E} \big[r^{def}|\pi^{def},\pi^{att}\big]$.

% \section{Background}

% To perform MCTS in NSGs, we design three neural networks in our method, which are a dynamic network for predicting transitions, a prior network for guiding exploration, and a value network for estimating the return of the leaf node in each simulation.

% {\color{red}The exploitability $\mathcal{E}_{def}$ of a defender policy $\pi_{def}$ is defined as the expected reward of the attacker given that the attacker best respond to $\pi_{def}$. Formally, $\mathcal{E}_{def}(\pi_{def})=\max_{\pi_{att}} \mathbb{E} \big[r_{att}|\pi_{def},\pi_{att}\big]$. The optimization objective of NSGs is to minimize the defender exploitability, i.e., $\min_{\pi_{def}}\mathcal{E}_{def}(\pi_{def})$, and it is equivalent to maximize the worst-case defender reward $\max_{\pi_{def}}\min_{\pi_{att}} \mathbb{E} \big[r_{def}|\pi_{def},\pi_{att}\big]$.}

\section{Efficiently Learning Non-Exploitable Policy}
In this section, we introduce our approach, NSGZero, for efficiently learning a non-exploitable policy in NSGs.
The algorithm improves data efficiency by performing planning with neural MCTS and improves scalability by enabling neural MCTS with decentralized execution.
We begin by introducing the DNNs required to perform neural MCTS in NSGs. Next, we introduce how NSGZero enables neural MCTS with decentralized execution. Finally, we present how to effectively train the DNNs in NSGZero.

\begin{figure*}
    \centering
    \includegraphics[width=1\textwidth]{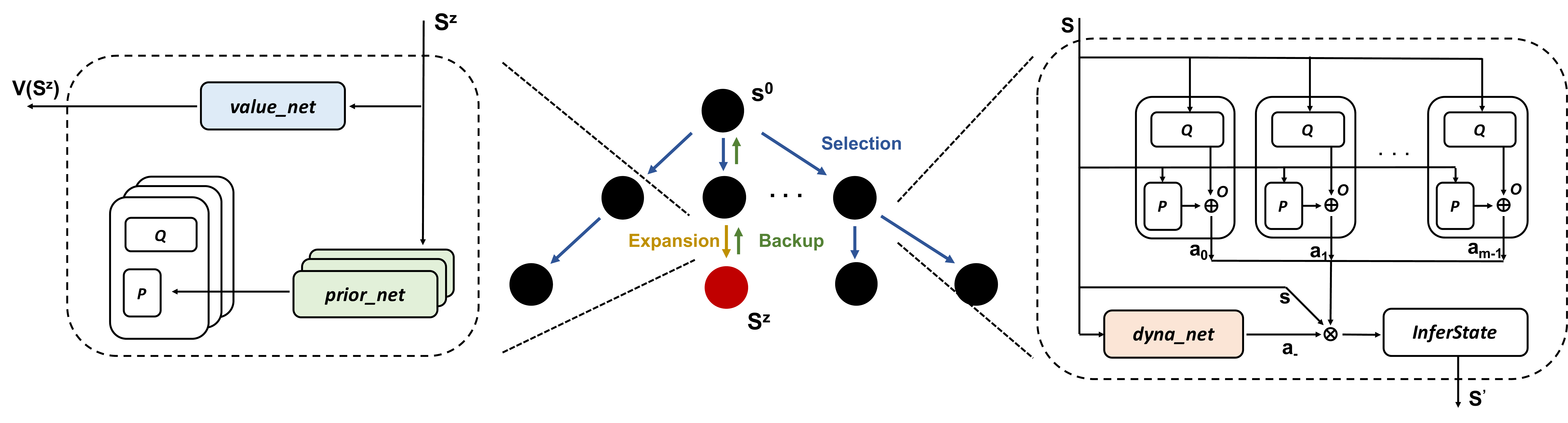}    \caption{\textbf{MCTS with the DNNs in NSGZero.} \textbf{Expansion}: When the search tree reaches a new state $s^z$, the prior network are invoked to predict prior policies for the resources, and the predicted polices are stored in $P(i,s,a)$. Meanwhile, the value network is applied to predict the state value $V(s^z)$, which is used to update $Q(i,s,a)$ in the backup phase. \textbf{Selection}: For each resource $i$, a hypothetical action $a_i$ is selected by comparing a score which is a weighted sum of $Q(i,s,a)$ and $P(i,s,a)$, and the weight is a function of $O(i,s,a)$. The dynamics network is used to predict the opponent's action $a_{-}$. With the hypothetical state $s$, the hypothetical actions for the $m$ resources $\boldsymbol{a}= \langle a_0, \dots, a_{m-1} \rangle$ and the predicted action for the opponent $a_{-}$, the next hypothetical state $s'$ can be inferred because the environment of NSGs is deterministic.}
    \label{fig.mcts_nn}
\end{figure*}
\subsection{Designing the DNNs Required by Neural MCTS}
To perform neural MCTS in NSGs, we should i) model the dynamics of the environment to perform state transition within the search tree; ii) predict the win rate (value) of a state to replace the expensive Monte Carlo rollout at a new state; and iii) leverage prior knowledge to guide exploration and narrow down the decision to high-probability actions. To achieve these purposes, we design three modules:

\textbf{Dynamics network.} In NSGs, the next state of a player depends heavily on what action is taken, because the action directly forms part of the state. Concretely, when the defender is in $s_t^{def}=\langle s_t^{att},l_t^{def}\rangle$ and takes action $a_t^{def}=\langle v_{t+1}^0,\dots, v_{t+1}^{m-1} \rangle$ (the movement of each resource), we can substitute $l_t^{def}$ with $a_t^{def}$ for constructing the next state $s_{t+1}^{def}=\langle s_{t+1}^{att},a_t^{def}\rangle$. The only unknown part in $s_{t+1}^{def}$ is $s_{t+1}^{att}=\langle s_t^{att}, v_{t+1}^{att}\rangle$, more specifically $v_{t+1}^{att}$ ($s_t^{att}$ is already known). Therefore, the dynamics network is designed to be a mapping from $s_t^{def}$ to $v_{t+1}^{att}$, to model the behavior of the opponent. Considering that legal actions change with states, we adopt a structure similiar to DRRN~\cite{acl} which learns representations for state and legal actions separately and outputs a likelihood by comparing the inner product of the representations. Formally, $dyna\_net(s,\mathcal{A}_{-}(s))=SoftMax(f(s)\odot g(\mathcal{A}_{-}(s)))$, where $\mathcal{A}_{-}(s)$ are the legal actions of the opponent at state $s$, $f$ and $g$ are feature extractors for state and action respectively, $\odot$ denotes inner product.
     
\textbf{Value network.} This module is designed for substituting the inefficient Monte Carlo rollout when estimating the value of a new state.
 Given a state, $value\_net: \mathcal{S} \to \mathbb{R}$ is able to predict the expected future return, which is equivalent to the expected end-game reward in NSGs.
      
\textbf{Prior network.} This module is used to incorporate prior knowledge into the selection phase of MCTS.
Specifically, the prior network takes state $s$ and legal actions $\mathcal{A}_i(s)$ (for resource $i$) as input and outputs a distribution over the actions, indicating the preferences for selecting each available action. Unlike the dynamics network and the value network that are shared among resources, the prior network is held by each resource individually. For each resource, the structure of its prior network is the same as that of the dynamics network, because these two types of DNNs are both used to predict the behavior of an agent in graphs. In practice, we apply parameter sharing among the prior networks, to transfer knowledge between the resources and accelerate training~\cite{gupta2017cooperative}.

Equipped with the three DNNs, NSGZero has the ability to simulate the future and plan based on the simulated state. 
At a decision point, NSGZero performs multiple simulations and generates policies based on the simulation results.

\subsection{Neural MCTS with Decentralized Execution}
Previous neural MCTS approaches, such as MuZero, focus on controlling a single agent, while in NSGs where the defender controls several resources, we need to enable neural MCTS with decentralized execution so that it can scale to the scenario with many resources. In this part, we introduce how to realize decentralized execution in NSGZero.
At each decision point, neural MCTS iterates $N$ simulations and outputs policies for resources based on the simulation results. We begin by discussing the simulation process. Then we introduce how to generate policies based on the simulation results.

\textbf{Simulation Process.} During the simulation process, for each resource $i$ and for each edge $(s,a)$ in the search tree $\Psi$, a set of statistics $\{Q(i,s,a), O(i,s,a), P(i,s,a)\}$ are stored, representing the state-action value, visit counts and prior policy respectively. Among the three types of statistics, $P(i,s,a)$ is calculated only once, at the step where the search tree reaches a new state. $Q(i,s,a)$ and $O(i,s,a)$ are continuously updated throughout the overall simulation process. As in Fig.~\ref{fig.mcts_nn}, each simulation consists of three phases: selection, expansion and backup:

\begin{itemize}[leftmargin=*]
\item \textbf{Selection:} In the selection phase, the search starts from the current hypothetical state $s^0$ and finishes when the simulation reaches a leaf node of the tree. To traverse within the search tree, NSGZero iteratively performs the selection operation. Assuming that it takes $z$ steps for NSGZero to reach a leaf node (in a simulation), for each hypothetical step $k=1, \dots, z$, each resource $i$ takes a hypothetical action $a^k_i$ according to the search policy:\begin{small}
\begin{multline}
\label{eq.select}
    a_i^k=\argmax_{a_i}\Big[
    C_{PUCT}\cdot \frac{\sqrt{\sum_{b_i} O(i,s^{k-1},b_i)}}{1+O(i,s^{k-1},a_i)} \cdot P(i,s^{k-1},a_i)\\
    +Q(i,s^{k-1},a_i)\Big]
\end{multline}
\end{small}where $O(i,s^{k-1},a_i)$ records the number of times resource $i$ takes action $a_i$ at hypothetical state $s^{k-1}$, $Q(i,s^{k-1},a_i)$ is the estimated value for action $a_i$ at hypothetical state $s^{k-1}$, $P(i,s^{k-1},a_i)$ is the prior probability of taking $a_i$ at hypothetical state $s^{k-1}$ which is calculated when the simulation first encounters hypothetical state $s^{k-1}$. Here, we use polynomial upper confidence trees (PUCT)~\cite{rosin2011multi,alphazero}, an adaptation of the standard MCTS, to combine value estimates with prior probabilities. A constant $C_{PUCT}$ is used to control the trade-off between $Q(i,s^{k-1},a_i)$ and $P(i,s^{k-1},a_i)$, i.e., exploitation and exploration. After selecting hypothetical actions $\boldsymbol{a}^k= \langle a_0^k, \dots, a_{m-1}^k \rangle$ for all resources, NSGZero predicts the behavior $a_{-}^k$ of the opponent by invoking the dynamics network. With the hypothetical state $s^{k-1}$, the hypothetical actions for the $m$ resources  $\boldsymbol{a}^k$ and the predicted action for the opponent $a_{-}^k$, we can infer the next hypothetical state $s^{k}$ and perform the next search process from $s^k$. 
    \item \textbf{Expansion:} The expansion phase starts when the transition $(s^{k-1}, \boldsymbol{a}^k, a_{-}^k)\to s^k$ leads to a state $s^z$ which is not in the search tree. In the expansion phase, NSGZero predicts the prior policy for each resource using the prior network and stores the probabilities in $P(i,s^k,a_i)$ correspondingly. In the meantime, since the search process has reached a node (state) which has never been visited before, NSGZero predicts the expected value $V(s^z)$ for the new state by calling the value network. The predicted state value is used to update $Q(i,s,a)$ along the trajectory from the root node to the parent of the leaf node.
    \item \textbf{Backup:} At the end of a simulation, the statistics, i.e., $Q(i,s^{k-1},a_i^{k})$ and $O(i,s^{k-1},a_i^{k})$, along the trajectory from the root node $s^0$ to the node $s^{z-1}$ are updated. For $k=z,\dots,1$, we estimate the cumulative discounted return $R^k$ by bootstrapping from $V(s^z)$. Formally, 
$
R^k=\sum_{\tau=0}^{z-k-1} \gamma^{\tau}\cdot r(s^{k+\tau},\boldsymbol{a}^{k+1+\tau},a_{-}^{k+1+\tau})+\gamma^{z-k}\cdot V(s^z)
$,
where $r(s^{k+\tau},\boldsymbol{a}^{k+1+\tau},a_{-}^{k+1+\tau})$ is the immediate reward for the transition $(s^{k-1}, \boldsymbol{a}^k, a_{-}^k)\to s^k$, $V(s^z)$ is the predicted state value for the hypothetical state $s^z$, and $\gamma$ is the discount factor. The estimated value $Q(i,s^{k-1},a_i^{k})$ and the visit counts $O(i,s^{k-1},a_i^{k})$ are updated as follows,
\begin{equation}
\begin{array}{l}
Q\left(i, s^{k-1}, a_i^{k}\right) \gets \frac{O\left(i, s^{k-1}, a_i^{k}\right) \cdot Q\left(i, s^{k-1}, a_i^{k}\right)+R^{k}}{O\left(i, s^{k-1}, a_i^{k}\right)+1}\\\\
O\left(i, s^{k-1}, a_i^{k}\right) \gets O\left(i, s^{k-1}, a_i^{k}\right)+1
\end{array}
\end{equation}
\end{itemize}
% \textbf{Backup:} At the end of a simulation, the statistics, i.e., $Q(i,s^{k-1},a_i^{k})$ and $O(i,s^{k-1},a_i^{k})$, along the trajectory from the root node $s^0$ to the node $s^{z-1}$ are updated. For $k=z\dots1$, we estimate the cumulative discounted return $R^k$ by bootstrapping from $V(s^z)$.
% Formally, 
% $
% R^k=\sum_{\tau=0}^{z-k-1} \gamma^{\tau}\cdot r(s^{k+\tau},\boldsymbol{a}^{k+1+\tau},a_{-}^{k+1+\tau})+\gamma^{z-k}\cdot V(s^z)
% $,
% where $r(s^{k+\tau},\boldsymbol{a}^{k+1+\tau},a_{-}^{k+1+\tau})$ is the immediate reward for the transition $(s^{k-1}, \boldsymbol{a}^k, a_{-}^k)\to s^k$, $\gamma$ is the discount factor. The estimated value $Q(i,s^{k-1},a_i^{k})$ and the visit counts $O(i,s^{k-1},a_i^{k})$ are updated as follows,
% \begin{equation}
% \begin{array}{l}
% Q\left(i, s^{k-1}, a_i^{k}\right)=\frac{O\left(i, s^{k-1}, a_i^{k}\right) \cdot Q\left(i, s^{k-1}, a_i^{k}\right)+R^{k}}{O\left(i, s^{k-1}, a_i^{k}\right)+1}\\\\
% O\left(i, s^{k-1}, a_i^{k}\right)=O\left(i, s^{k-1}, a_i^{k}\right)+1
% \end{array}
% \end{equation}
An overview about how to perform search in a simulation is presented in Algo.~\ref{algo.search}. 

\textbf{Generating the Policies.} At each decision point $s_t$, NSGZero first initializes the search tree with $s_t$ as the unique node and all former statistics being cleared. Then it repeatedly performs $N$ simulations, gradually growing the search tree and tracking the statistics. Note that $s_t$ denotes a real state, as contrast to $s^k$ which is the $k$-th hypothetical state. A simulation starts by performing the selection operation from $s_t$ and it takes $s_t$ as the $0$-th hypothetical state ($s^0 \leftarrow s_t$).
After the $N$ simulations, NSGZero generates a policy for each resource by querying the visit frequency of legal actions, with a temperature parameter $\mathcal{T}$ to control the randomness of the distribution. Formally, the policy for resource $i$ is $\pi_i(s_t,a_i) \propto \frac{O(i, s_t,a_i)^{1/\mathcal{T}}}{\sum_{b_i} O(i, s_t, b_i)^{1/\mathcal{T}}}$. Each resource $i$ samples its action $a_{i,t}$ from $\pi_i(s_t)$ respectively.
The pseudo-code of the overall execution process is in Algo.~\ref{algo.exe}.
\begin{algorithm}[t]
\caption{NGSZero-$\Psi$.SEARCH}
\label{algo.search}
\KwIn{A hypothetical state $s$.}
\uIf{
$s$ is an ending state}{\Return the ending-game reward $r(s)$\;}
\uElseIf{$s$ is not in the search tree $\Psi$}
{Add $s$ to the search tree $\Psi$\;
\For{resources $i=0, \dots, m-1$}{
$\Psi.P(i,s)$$\leftarrow$$\Psi.prior\_net(s,\mathcal{A}_i(s))$\;
}
\Return the predicted reward $\Psi.value\_net(s)$\;
}
\Else{
\For{resources $i=0, \dots, m-1$}{
Select hypothetical action $a_i$ (Eq.~\ref{eq.select})\;
}
$a_{-}\sim\Psi.dyna\_net(s,\mathcal{A}_{-}(s))$ $\backslash\backslash$ the opponent\; 
$s'$ $\leftarrow$ InferState(s, $\boldsymbol{a}$, $a_-$), $\boldsymbol{a}= \langle a_0, \dots, a_{m-1} \rangle$\;
$r(s')\leftarrow\Psi.search(s')$\;
\For{resources $i=0, \dots, m-1$}{
Update $\Psi.Q(i,s,a_i)$ and $\Psi.O(i,s,a_i)$\;
}
\Return the searched reward $r(s').$
}
\end{algorithm}

\begin{algorithm}[t]
\caption{NGSZero-EXECUTION}
\label{algo.exe}
\KwIn{The current state $s_t$, the search tree $\Psi$.}
$\Psi.clear()$  ~$\backslash\backslash$ clear statistics stored in the search tree\;
\For{$N$ simulations}{
$\Psi.search(s_t)$ ~$\backslash\backslash$ perform lookahead search\;
}
\For{resources $i=0, \dots, m-1$}{
$\pi_i(s_t,a_i) \propto \frac{O(i, s_t,a_i)^{1/\mathcal{T}}}{\sum_{b_i} O(i, s_t, b_i)^{1/\mathcal{T}}}$; ~~$a_{i,t} \sim \pi_i(s_t)$\;
}
\KwOut{Joint action $\boldsymbol{a}_t= \langle a_{0,t}, \dots, a_{m-1,t} \rangle$.}
\end{algorithm}

\subsection{Training in NSGZero}
All parameters of the DNNs in NSGZero are trained jointly to match corresponding targets collected by repeatedly playing NSGs. For brevity, we use $\Psi_p$, $\Psi_v$, $\Psi_d$ to denote the prior network, the value network and the dynamics network respectively.
Our first objective is to minimize the error between the prior policies predicted by $\Psi_p$ and the improved policies $\boldsymbol{\pi}(s)= \langle \pi_0(s), \dots, \pi_{m-1}(s) \rangle$ searched by NSGZero. To share experiences between resources and simplify the networks to optimize, we propose to let all resources share the parameters of their prior networks. As in Algo.~\ref{algo.exe} (lines 5-7), NSGZero generates a policy $\pi_i(s)$ and samples an action $a_i$ for each resource $i$. Therefore, we update $\Psi_p$ by minimizing $-\frac{1}{m}\sum_{i=0}^{m-1} a_i \cdot \log (\Psi_p(s, \mathcal{A}_i(s)))$, here $a_i$ is in one-hot form. Next, the value network $\Psi_v$ needs to be optimized to match discounted return. Let $h$ be the length (total time steps) of an episode and $r$ be the end-game reward, for $1\leq t\leq h$, the discounted return is $\gamma^{h-t}\cdot r$. The conventional optimization method for value function is to minimize the mean squared error (MSE)~\cite{dqn}, i.e., minimizing $(\Psi_v(s)-\gamma^{h-t}\cdot r)^2$. However, in NSGs, the target is bounded within the $[0,1]$ interval, so we propose to convert the optimization of $\Psi_v$ to a classification problem, i.e., optimize it by minimizing the binary cross entropy (CE) loss, with $\gamma^{h-t}\cdot r$ as soft label. Formally, $l_v=-(\gamma^{h-t}\cdot r)\cdot \log(\Psi_v(s))-(1-\gamma^{h-t}\cdot r)\cdot \log(1-\Psi_v(s))$. The optimization of the dynamics network $\Psi_d$ is similar to that of $\Psi_p$, with the objective as $-a_{-} \cdot \log (\Psi_d(s, \mathcal{A}_{-}(s)))$, and $a_{-}$ is a one-hot label, indicating the opponent's action. The overall loss is
% \begin{small}
\begin{multline}
   \mathcal{L}_{p,v,d}=-\frac{1}{B}\sum_{b=0}^{B-1}\frac{1}{h_b}\sum_{t=1}^{h_b}\Big[ \frac{1}{m}\sum_{i=0}^{m-1} a_{i,t} \cdot \log (\Psi_p(s_t, \mathcal{A}_i(s_t))\\
   +(1-\gamma^{h_b-t}\cdot r_b)\cdot \log(1-\Psi_v(s_t))\\
   +(\gamma^{h_b-t}\cdot r_b)\cdot \log(\Psi_v(s_t))\\
    +a_{-,t} \cdot \log (\Psi_d(s_t, \mathcal{A}_{-}(s_t)))\Big]
\end{multline}
where $B$ is the batch size, $h_b$ and $r_b$ denote the length and the end-game reward of an episode $b$.

\textbf{Implementation.} Optimizing $\mathcal{L}_{p,v,d}$ with batch training is non-trivial because the length of each episode varies. In practice, for each episode $b$, we store the trajectories of the defender and the attacker, then we can infer all the variables required to calculate the loss for this episode, i.e., $r_b$, $h_b$, and $s_t$. For a batch of episodes $B$, we pad all the trajectories within the batch to the same length as the time horizon $T$ (pad with 0). Then we calculate the loss in a batch for the $B$ episodes by iterating over the time steps. For the padded entries, they do not contribute to the overall loss, so the calculated values for them should be filtered out. We generate a mask with $h_b$ for each episode $b$, to indicate the padding items, then use the mask to filter out the values of padded items. Finally, the overall loss $\mathcal{L}_{p,v,d}$ is calculated by averaging all valid entries within the batch.

\textbf{Modeling the Attacker.} Despite self-play MCTS having been widely used in symmetric games~\cite{silver2016mastering}, in NSGs which are asymmetric, we need to model the defender and the attacker differently. We apply high-level actions to the attacker to achieve efficient training~\cite{xue2021solving}. Specifically, the attacker makes decisions on which target to be attacked, rather than deciding where to go in the next time step. At the beginning of each episode, the attacker selects a target and samples a path to the chosen target, then he moves along this path at each step. We use Multi-Armed Bandit (MAB), an algorithm to optimize the decision of multiple actions, to estimate the value of each target according to the latest $J$ plays. Formally, the estimated value for a target is $Q(\zeta)=\frac{\sum_{j=1}^J r_j\cdot\mathbb{I}[\zeta_j=\zeta]}{\sum_{j=1}^J\mathbb{I}[\zeta_j=\zeta]}$, where $\zeta$ denotes the target, $r_j$ is the player’s reward for the $j$-th episode, and $\mathbb{I}$ is the binary indicator function. The MAB acts by selecting $\zeta$ with the largest estimated value. There is an Averager (AVGer) that tracks the historical behaviour of the MAB, by counting the number of times each target has been selected by the MAB. The AVGer generates a categorical distribution according to the counts and makes decision by sampling from the distribution. The attacker behaves as a mixture of the MAB and the AVGer, with an anticipate parameter $\eta$ indicating the probability of him following the MAB.
\begin{table*}
\centering
\scalebox{1}{
\begin{tabular}{c|c|c|c|c}
\hline 
 & \multirow{1}{*}{\begin{tabular}[c]{@{}c@{}}Manhattan: $m=3$ \end{tabular}} & \multirow{1}{*}{\begin{tabular}[c]{@{}c@{}}Manhattan: $m=6$\end{tabular}} & \multirow{1}{*}{\begin{tabular}[c]{@{}c@{}}Singapore: $m=4$\end{tabular}}&\multirow{1}{*}{\begin{tabular}[c]{@{}c@{}}Singapore: $m=8$ \end{tabular}}\\ 
% &&&&\\
   \hline\hline
NSGZero (Ours)  & \textbf{0.1670 $\pm$ 0.0231}   & \textbf{0.3660 $\pm$ 0.0299}  & \textbf{0.1810 $\pm$ 0.0239}   & \textbf{0.4140$\pm$ 0.0306}   \\ \hline
NSG-NFSP~\cite{xue2021solving} & 0.0390 $\pm$ 0.0120  &      $-$     & 0.0130 $\pm$ 0.0070   & $-$    \\ \hline 
IGRS++~\cite{yz19} & OOM  &      OOM    & OOM  & OOM      \\ \hline 
Uniform 
 Policy &  0.0120 $\pm$ 0.0096   &      0.1020 $\pm$ 0.0188  & 0.0240 $\pm$0.0095   &  0.2590 $\pm$ 0.0272         \\ \hline 
\end{tabular}
}
\caption{Approximate worst-case defender rewards, averaged over 1000 test episodes. OOM stands for Out of Memory. The ``$\pm$'' indicates $95\%$
    confidence intervals over the 1000 plays.}
\label{tab.def}
\end{table*}

\begin{figure}
    \centering
    \includegraphics[width=0.47\textwidth]{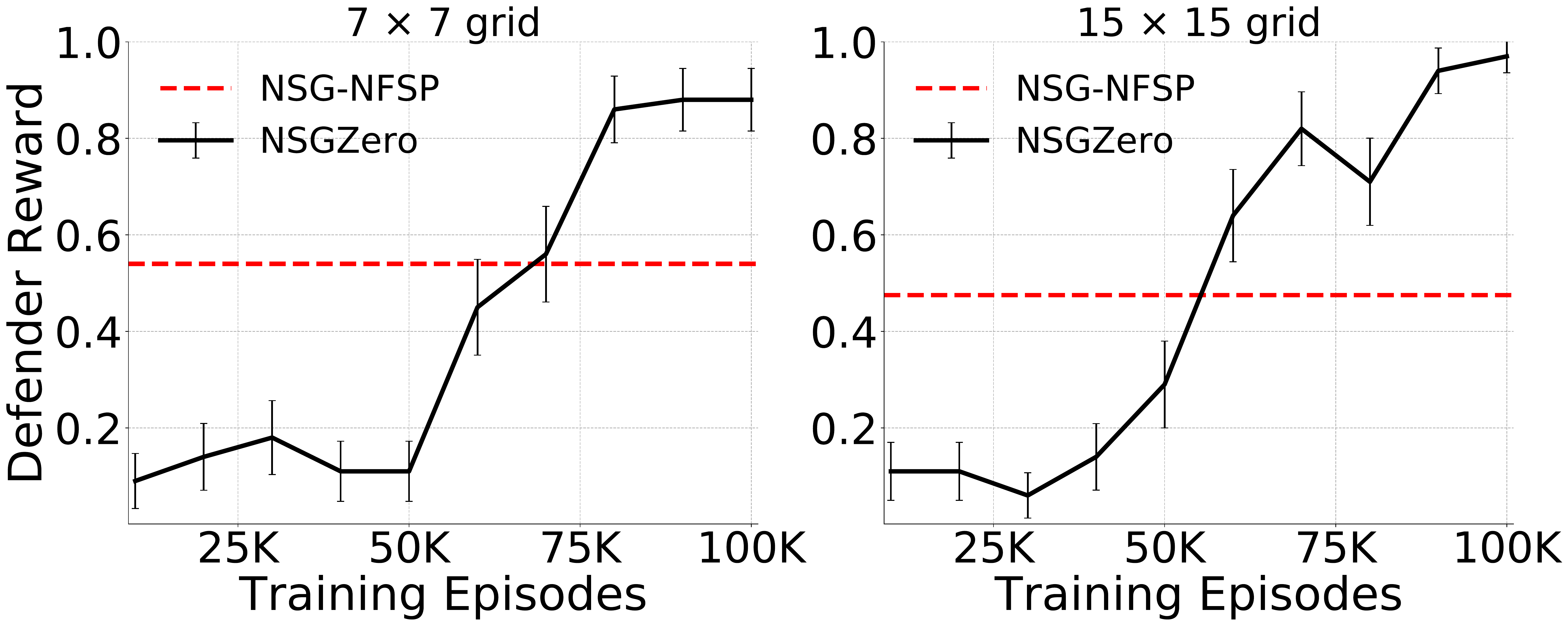}
    \caption{The worst-case defender reward. Error bars indicate 95\% confidence intervals over the 100 testing episodes. The red dashed line indicates the performance of NSG-NFSP after training 1 million episodes.}
    \label{fig.data}
\end{figure}

\section{Experiments}
We evaluated the performance of NSGZero on a variety
of NSGs with different scales. We use these experiments to answer three questions. First, whether NSGZero is sufficiently data efficient compared to state-of-the-art algorithms, i.e., whether the algorithms can achieve comparable or even better performance when learning from fewer experiences. Second, whether the algorithm has better scalability and is applicable to large-scale NSGs. Third, how components of NSGZero affect the performance and whether the DNNs are trained as we expect. Experiments are conducted on a server with a 20-core 2.10GHz Intel Xeon Gold 5218R CPU and an NVIDIA RTX 3090 GPU.
\subsection{Data Efficiency}
To answer the first question, i.e., whether NSGZero has better data efficiency compared to state-of-the-art learning approaches, we evaluate the performance of NSGZero on two NSGs with synthetic graphs. For the first NSG, its graph is a $7\times7$ grid, with vertical/horizontal edges appearing with probability 0.5 and diagonal edges appearing with probability 0.1. We initialize the location of the attacker at the center of the grid and let resources of $m=4$ be located uniformly around the attacker. There are 10 target nodes distributed randomly at the boundary of the grid. The time horizon is set to be equal to the length of the grid, i.e., 7. For the second NSG, it is generated similarly except that the graph is a randomly sampled $15\times15$ grid (vertical/horizontal edges appear with probability 0.4 and diagonal edges appear with probability 0.1) and the time horizon is 15. We find the best response attacker by enumerating all attack paths and choosing the path which leads to the lowest defender reward. The worst-case defender reward is the defender's payoff when she plays against the best response attacker.

We train NSGZero for 100,000 episodes and plot the worst-case defender reward. As in Fig.~\ref{fig.data}, the worst-case defender reward has seen a stable increase as training proceeds and it can reach values at around 0.88 and 0.95 for the $7\times7$ grid and the $15\times15$ grid, respectively. Considering that the worst-case defender reward is upper-bounded by 1, NSGZero has found a near-optimal solution for both NSGs. Comparisons are made with a state-of-the-art learning method NSG-NFSP. We first train NSG-NFSP for 100,000 episodes, but find that there is no obvious learning effect and the worst-case defender reward remains close 0. Therefore, we increase its training episodes by 10 times to 1 million. Then NSG-NFSP reaches values of 0.54 and 0.48 for the two NSGs, still significantly lower than that of NSGZero. The experiments show that NSGZero is significantly more data efficient than NSG-NFSP, and it is able to achieve better performance even when learning from fewer experiences. 

\begin{figure*} 
    \centering
     \begin{subfigure}[t]{0.33\textwidth}
         \centering
         \includegraphics[width=1\textwidth]{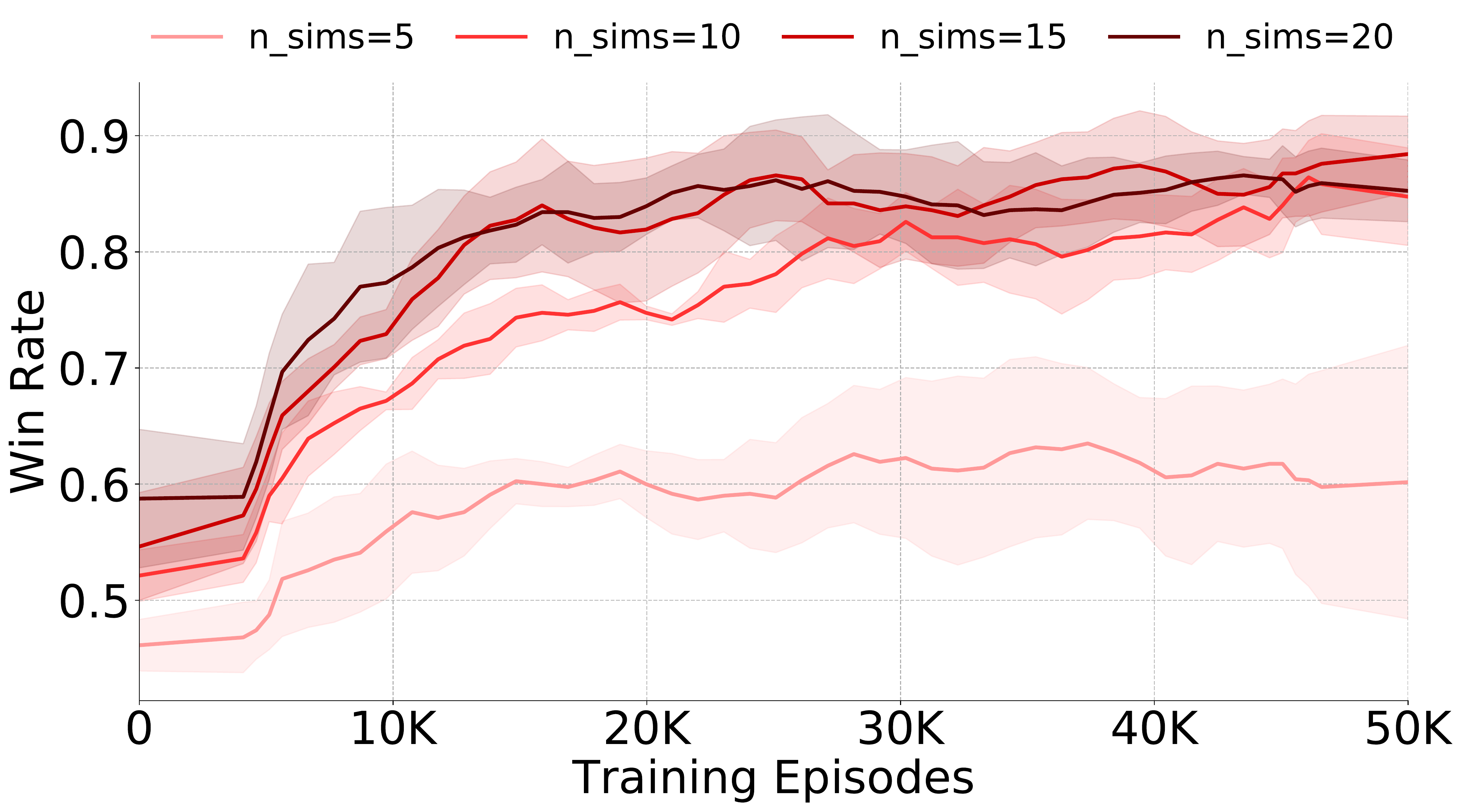}
         \caption{Number of simulations}
         \label{fig.n_sims}
     \end{subfigure}
     \hfill
     \begin{subfigure}[t]{0.33\textwidth}
         \centering
         \includegraphics[width=1\textwidth]{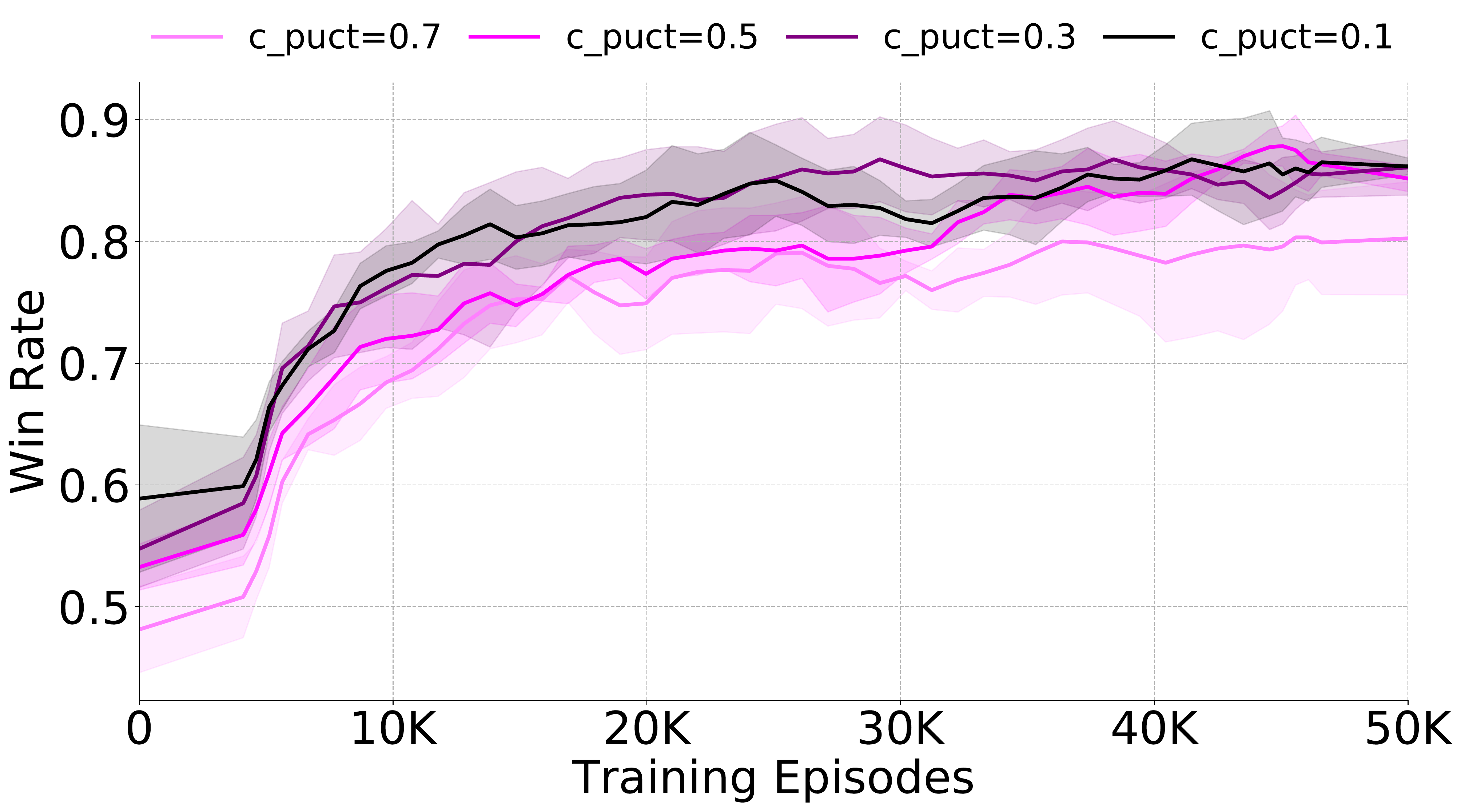}
         \caption{Exploration constant}
         \label{fig.cpuct}
     \end{subfigure}
     \begin{subfigure}[t]{0.33\textwidth}
         \centering
         \includegraphics[width=1\textwidth]{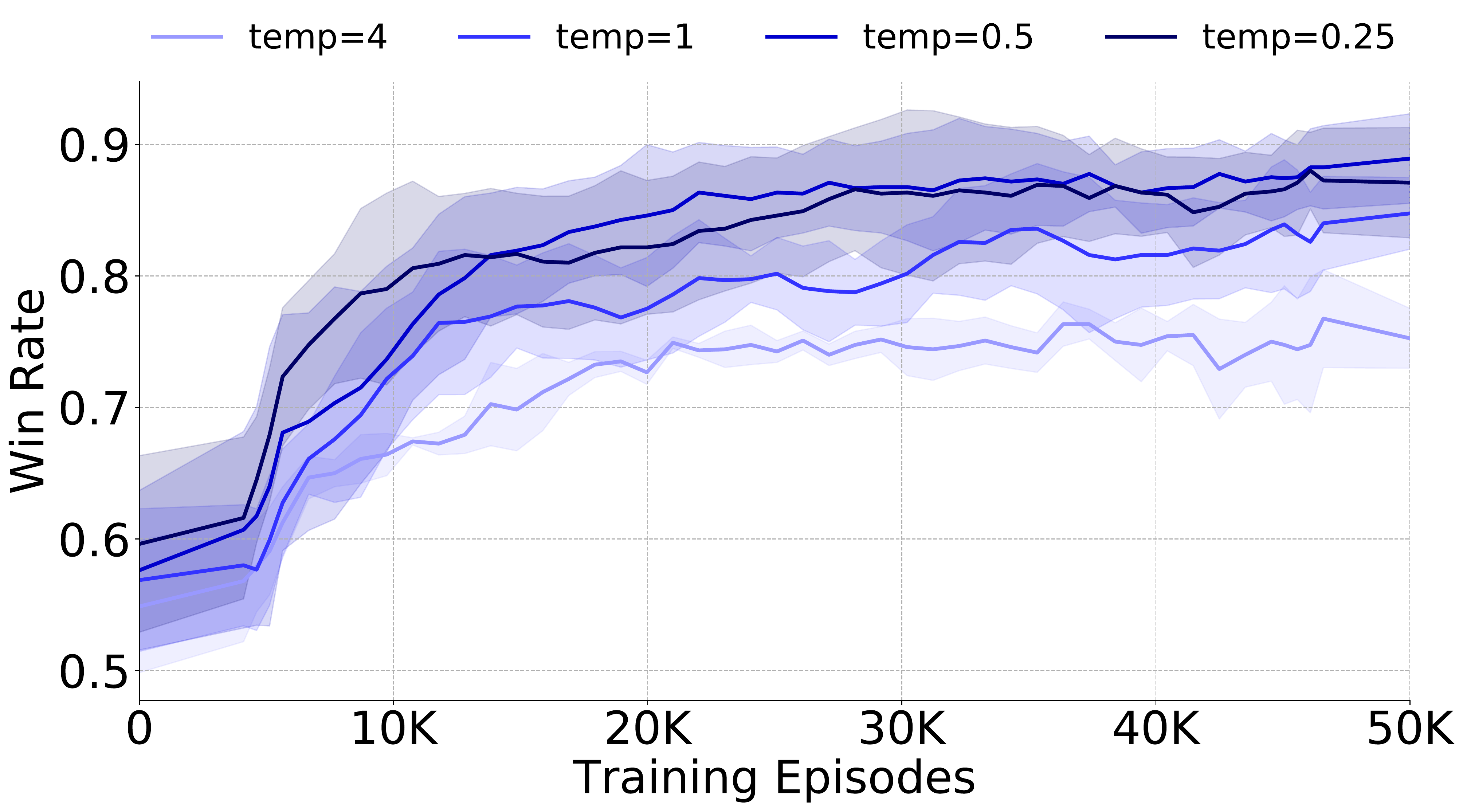}
         \caption{Temperature parameter}
         \label{fig.temp}
     \end{subfigure}
     \caption{Ablation studies for hyper-parameters in the execution process.
     The learning curves are for the NSGZero defender which plays against a heuristic uniform attacker, averaged over 5 runs.}
      \label{seq}
\end{figure*}
\subsection{Scalability of NSGZero}
To examine whether NSGZero is more scalable compared to existing methods, we extract two real-world maps from Manhattan and Singapore vis OSMnx~\cite{osmnx} and create two large-scale NSGs for each map. For the Manhattan map, the first NSG has 3 resources and 7 targets. In the second NSG, we increase the scale to 6 resources and 9 targets to evaluate the performance of NSGZero in an NSG with many secure resources. For the Singapore map, the setup is similar, with the first NSG having 4 resources and 10 targets and the second NSG having 8 resources and 14 targets. Time horizon $T$ is set as 30 for all NSGs to ensure that the policy space is sufficiently large.
We manually set the initial location of the attacker, the resources and the targets to make the NSGs more realistic (details are in the appendix). When performing evaluation, considering that we cannot enumerate all attack paths as is done in small NSGs, as a mitigation, we use all the shortest paths and a best response DQN attacker to approximate the worst case. The smallest defender reward for playing against the worst-case approximators is reported. 

We make comparisons with two state-of-art-algorithms, i.e., NSG-NFSP~\cite{xue2021solving} and IGRS++~\cite{yz19}, and a heuristic uniform policy. For the two learning-based approaches, i.e., NSGZero and NSG-NFSP, we train the model for 100,000 episodes. As shown in Table~\ref{tab.def}, our algorithm significantly outperforms the baselines. IGRS++, as an incremental strategy generation algorithm, requires that all attack paths to be enumerable, otherwise it runs out of memory due to lack of a terminal state. However, this requirement is not satisfied in the created games. IGRS++ fails to solve the games. For NSG-NFSP, when $m$ is small, it can learn a policy, but the performance is not satisfactory, just comparable to uniform policy. We infer the reason is because of data inefficiency, i.e., NSG-NFSP is unable to learn a good policy from just 100,000 episodes. When $m$ of the two maps is increased to 6 and 8, respectively, the training of NSG-NFSP becomes infeasible due to the huge action space. NSGZero performs well in all four NSGs, and it can scale to NSGs with many security resources, which are unsolvable by existing methods. 
\begin{figure}[t]
    \centering
    \includegraphics[width=0.4\textwidth]{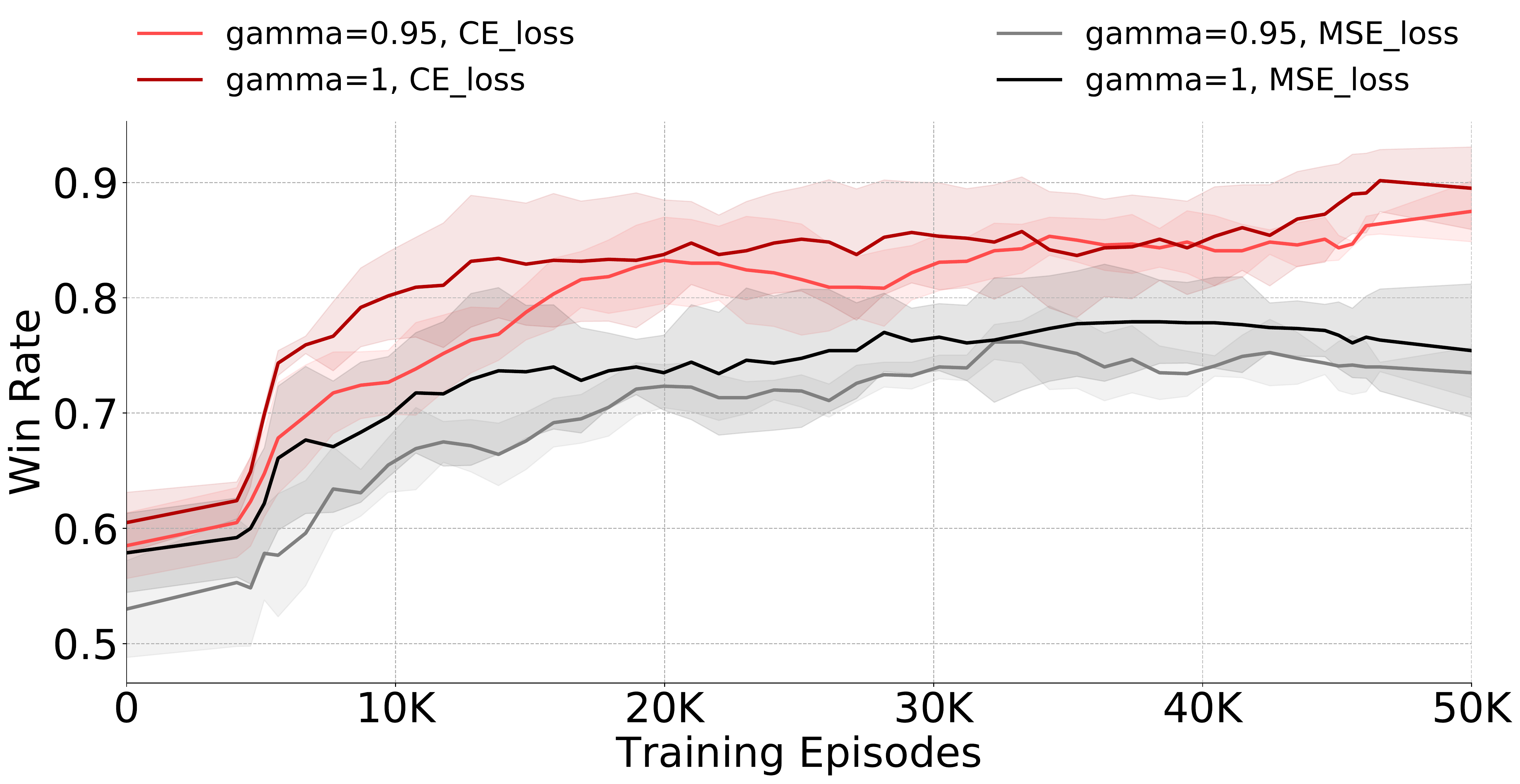}
    \caption{Ablation studies for the training process.
     The learning curves are for the NSGZero defender which plays against a heuristic uniform attacker, averaged over 5 runs.}
    \label{fig.ab_loss}
\end{figure}
\begin{figure}
    \centering
    \includegraphics[width=0.47\textwidth]{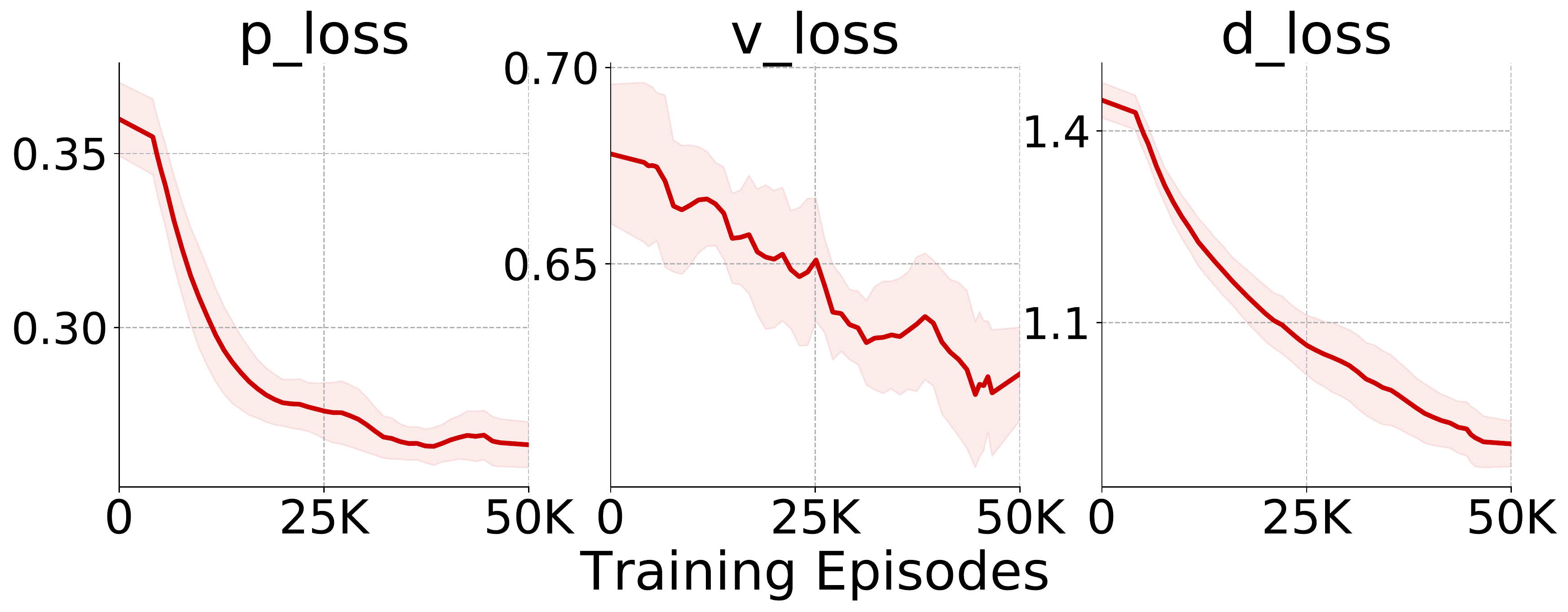}
    \caption{The loss curves of the prior network, the value network, and the dynamics network, averaged over 5 runs.}
    \label{fig.loss}
\end{figure}

\subsection{Ablation Study and Analysis}
To investigate how each component affects NSGZero, we evaluate the performance of NSGZero under different hyper-parameters, which leads to different structures of search trees or different policy generating strategies. We create an NSG by randomly sampling a $7\times7$ grid, and the setup is similar as before, i.e., vertical/horizontal edges appear with probability 0.5 and diagonal edges appear with probability 0.1. There are $m=4$ resources and 10 targets. The time horizon is 7.
% with vertical/horizontal edges appearing with probability 0.5 and diagonal edges appearing with probability 0.1. We initialize the location of the attacker at the center of the grid and let $m=4$ resources locate uniformly around the attacker. There are 10 target nodes distributed randomly at the boundary of the grid. The time horizon is set to be equal to the length of the grid. 
For the attacker, he uses a fixed strategy: at the beginning of each episode, he draws a target as a destination and randomly samples a path to the chosen target. At each step of the episode, the attacker moves according to the path. We use NSGZero to model the defender.

\textbf{Number of Simulations.} NSGZero generates policies according to the search tree, which starts from a single root node and gradually expands by performing $N$ simulations. Therefore, we first examine how the number of simulations $N$ affects the performance of NSGZero. We set $N$ to 5, 10, 15, and 20, respectively, and test the average win rate of the defender throughout the training process. As in Fig.~\ref{fig.n_sims}, the performance of NSGZero becomes better in general as $N$ increases. When $N=5$, the performance is unsatisfactory and the win rate increases from 0.47 to around 0.6. As we increase the number of simulations, obvious learning effect can be observed, with the win rate reaching around 0.9. We also find that $N=15$ is sufficient for good performance, after which increasing the number of simulations is no longer beneficial.

\textbf{Exploration Constant.} NSGZero uses an exploration constant $C_{PUCT}$ to control the extent of exploration when performing searches. To study how this parameter affects the performance, we set $C_{PUCT}\in\{0.7,0.5,0.3,0.1\}$. Fig.~\ref{fig.cpuct} shows that a large $C_{PUCT}$ could hurt the performance since it weights more on the prior knowledge and the exploration term may overwhelm the value estimation. We can also find that a small $C_{PUCT}$ is sufficient to narrow down the selection to high-probability actions.

\textbf{Temperature Parameter.} The number of simulations $N$ and the exploration constant $C_{PUCT}$ influence the simulation process, while after the simulations, NSGZero generates policies by counting the visit frequency and this procedure is controlled by the temperature parameter. To investigate the effect of the temperature parameter on NSGZero, we set the temperature to 4, 1, 0.5 and 0.25 to introduce different degrees of randomness when generating the policies. As in Fig.~\ref{fig.temp}, high temperature causes a decrease in performance. When the temperature equals to 4, the final win rate is about 0.75, significantly lower than the win rate of 0.9 at lower temperatures. Meanwhile, low temperature can also lead to a slight decrease in win rate, in which case NSGZero is too confident of the generated policies and does not introduce enough randomness.

Previous experiments investigated three factors affecting the execution phase of NGSZero. Next, we examine those factors that influence the training process. In NGSZero, when training the value network, we convert its optimization objective from Mean Square Error (MSE) loss to binary cross entropy (CE) loss. To examine whether this change is beneficial, we plot the learning curves for these two optimization objectives respectively. As shown in Fig.~\ref{fig.ab_loss}, CE loss significantly outperforms MSE loss throughout the training, with a final win rate of 0.9 compared to a value of only about 0.75 for the MSE loss. We further investigate the effect of the discount factor $\gamma$, which influences the optimization target of the value network ($\gamma^{h-t}\cdot r$). We find that, in NSGs where only end-game rewards are available, not applying a discount to rewards can result in consistently better performance. As in Fig.~\ref{fig.ab_loss}, $\gamma=1$ outperforms $\gamma=0.95$ for both MSE loss and CE loss.
The reason may be that the distribution of the target is easier to learn when the rewards are not discounted. 
To examine whether the three DNNs of NSGZero are trained to match their corresponding targets, we plot the loss curves for them respectively. As shown in Fig.~\ref{fig.loss}, significant reduction in loss can be found in all the DNNs, indicating that the parameters of the DNNs are optimized properly.
\section{Conclusion}
In this paper, we introduce NSGZero as a learning method to efficiently approach a non-exploitable policy in NSGs. We design three DNNs, i.e., the dynamics network, the value network and the prior network, to unlock the use to efficient MCTS in NSGs. To improve scalability, we enable decentralized control in neural MCTS, which makes NSGZero applicable to NSGs with a large number of resources. To optimize the parameters of the DNNs, we provide a learning paradigm to achieve effective joint training in NSGZero. Experiments are conducted on a variety of NSGs with different graphs and scales. Empirical results show that, compared to state-of-the-art algorithms, NSGZero is significantly more data efficient and it is able to achieve much better performance even when learning from fewer experiences. Furthermore, NSGZero can scale to large-scale NSGs which is unsolvable by existing approaches.

\section*{Acknowledgements}
This research is partially supported by Singtel Cognitive and Artificial Intelligence Lab for Enterprises (SCALE@NTU), which is a collaboration between Singapore Telecommunications Limited (Singtel) and Nanyang Technological University (NTU) that is funded by the Singapore Government through the Industry Alignment Fund – Industry Collaboration Projects Grant.
% to form the search tree of NSGZero and present how NSGZero performs efficient lookahead decision. We propose an effective learning paradigm to optimize the DNNs in NSGZero. Experiments show that, compared to existing algorithms, our method has significant better data efficiency and scalability.

% \clearpage
% \bibliographystyle{named}
\bibliography{aaai22}

\newpage
\appendix
\onecolumn

\begin{figure*}
    \centering
     \begin{subfigure}[t]{0.6\textwidth}
         \centering
         \includegraphics[width=1\textwidth]{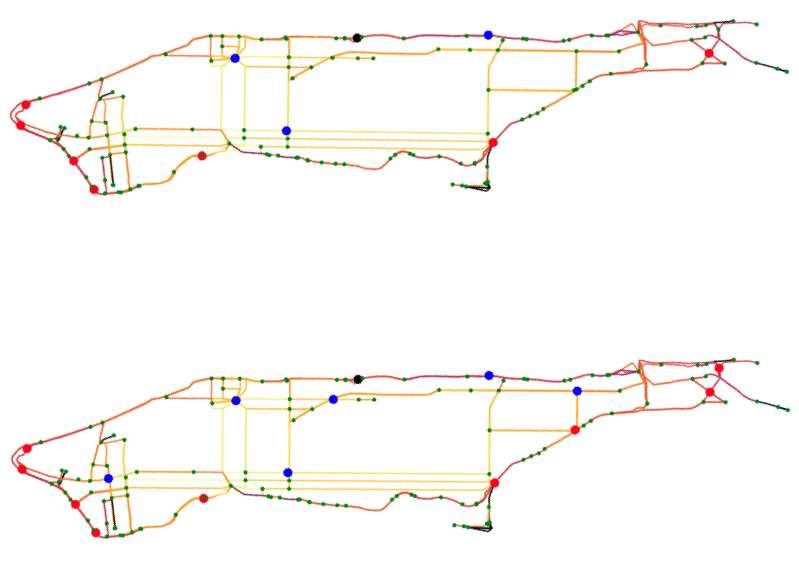}
         \caption{Manhattan}
     \end{subfigure}
     \vfill
     \vspace{1cm}
     \begin{subfigure}[t]{0.52\textwidth}
         \centering
         \includegraphics[width=1\textwidth]{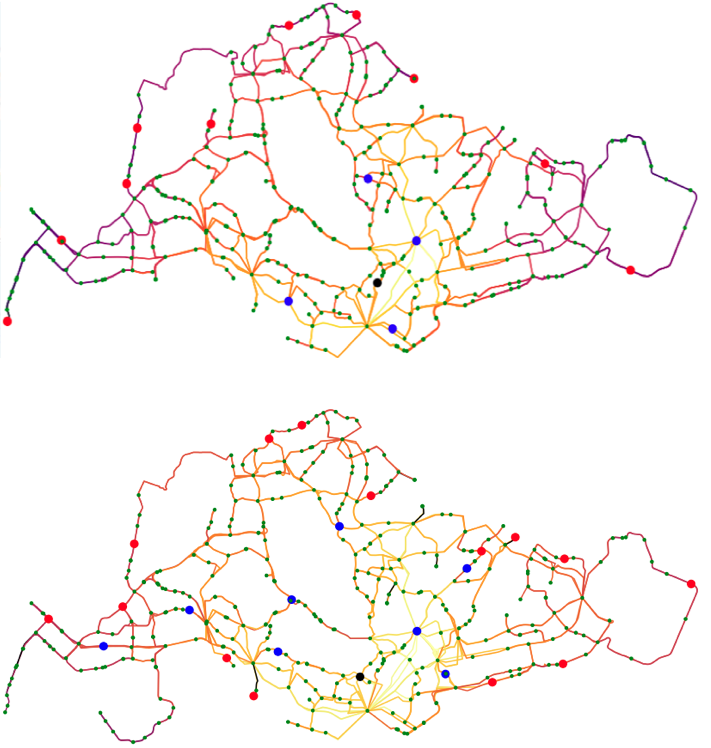}
         \vspace{0.2cm}
         \caption{Singapore}
     \end{subfigure}
\caption{The extracted real-world maps. The attacker, the resources and the targets are marked by dark point, blue points and red points, respectively.}
\end{figure*}
\end{document}